\documentclass[letterpaper, 10pt, conference]{ieeeconf}
\IEEEoverridecommandlockouts
\usepackage{cite}
\usepackage{amsmath,amssymb,amsfonts}
\usepackage{algorithmic}
\usepackage{graphicx}
\usepackage{textcomp}
\usepackage{xcolor}
\usepackage{dirtytalk}
\usepackage{comment}
\usepackage{float}
\newcommand{\ec}{ {\bf\color{red}{EY:} }}
\def\BibTeX{{\rm B\kern-.05em{\sc i\kern-.025em b}\kern-.08em
    T\kern-.1667em\lower.7ex\hbox{E}\kern-.125emX}}
\newtheorem{assumption}{Assumption}

\title{%\LARGE \bf
Optimal reporter placement in sparsely measured genetic networks using the Koopman operator
}

\author{Aqib Hasnain, Nibodh Boddupalli, and Enoch Yeung% <-this % stops a space
\thanks{A. Hasnain and N. Boddupalli are with the Department of Mechanical Engineering, University of California, Santa Barbara  {\tt\small aqib@ucsb.edu}, {\tt\small nibodh@ucsb.edu} \newline \indent E. Yeung is with the Department of Mechanical Engineering, Center for Control, Dynamical Systems, and Computation, and
Biomolecular Science and Engineering, University of California, Santa Barbara {\tt\small eyeung@ucsb.edu}}
}

\begin{document}

% \title{Optimal reporter placement in sparsely measured genetic networks using the Koopman operator  \\
% %{\footnotesize \textsuperscript{*}Note: Sub-titles are not captured in Xplore and
% %should not be used}
% \thanks{A. Hasnain and N. Boddupalli are with the Department of Mechanical Engineering, University of California, Santa Barbara  {\tt\small aqib@ucsb.edu}, {\tt\small nibodh@ucsb.edu} \newline \indent E. Yeung is with the Department of Mechanical Engineering, Center for Control, Dynamical Systems, and Computation, and
% Biomolecular Science and Engineering, University of California, Santa Barbara {\tt\small eyeung@ucsb.edu}}
% }
% \author{\IEEEauthorblockN{Aqib Hasnain, Nibodh Boddupalli, and Enoch Yeung}

% }

\maketitle

\begin{abstract}
Optimal sensor placement is an important yet unsolved problem in control theory. In biological organisms, genetic activity is often highly nonlinear, making it difficult to design libraries of promoters to act as reporters of the cell state. We make use of the Koopman observability gramian to develop an algorithm for optimal sensor (or reporter) placement for discrete time nonlinear dynamical systems to ease the difficulty of design of the promoter library. This ease is enabled due to the fact that the Koopman operator represents the evolution of a nonlinear system linearly by lifting the states to an infinite-dimensional space of observables. The Koopman framework ideally demands high temporal resolution, but data in biology are often sampled sparsely in time. Therefore we compute what we call the temporally \textit{fine-grained} Koopman operator from the temporally \textit{coarse-grained} Koopman operator, the latter of which is identified from the sparse data. The optimal placement of sensors then corresponds to maximizing the observability of the fine-grained system. We demonstrate the algorithm on a simulation example of a circadian oscillator.  
\end{abstract}

%\begin{IEEEkeywords}
%component, formatting, style, styling, insert
%\end{IEEEkeywords}

\section{Introduction}

Spectral methods have been increasingly popular in data-driven analysis of nonlinear dynamical systems. Recently, researchers working in Koopman operator theory have shown that it is possible to identify and learn the fundamental modes for a nonlinear dynamical system from data \cite{rowleySpectral2009,proctor_brunton_kutz_2016}. 
This operator, originally defined nearly 100 years ago by Koopman \cite{koopman}, is a linear infinite dimensional operator that fully describes the underlying nonlinear dynamical system. Identifying Koopman operators from data has become
computationally tractable, largely due to advances integrating machine learning and deep learning to generate novel, efficient representations of observable subspaces for the Koopman operator \cite{yeung2017learning,lusch2018deep}.

In many high-dimensional nonlinear systems, typically it is not physically or economically feasible to measure every state with the resolution specified by a fine-grained temporal model. For example, the bacteria \textit{E. coli} have approximately 4400 genes, making both spatially and temporally fine data collection nearly impossible. On one hand, high-coverage omics measurements provide a system-level view of all gene activity, but prohibitive costs and the laborious and destructive nature of sampling make it difficult to resolve dynamics at a high temporal resolution.  On the other hand, fluorescently tagged genes can be measured at the second to minutes timescale, to profile bursty RNA dynamics and protein expression.  Knowing which genes to tag with fluorescent markers is critical, since not every gene can be simultaneously tagged. This challenge motivates the need for algorithmic data-driven approaches which allow the user (e.g. biologists) to know \textit{a priori} which genes should be sampled.  Finally, is it possible to design a nonlinear observer that rather than measuring a single gene or a single node in the network, fuses the state of a select set of biomarker genes to report out an aggregate cellular state of the system?  The fundamental question is how to use metrics for nonlinear observability to design observers or optimize sensor placement. Sinha et al. presented a systematic framework based on linear transfer operators for the optimal placement of sensors and actuators for control of nonequilibrium dynamics \cite{sinha2016operator}. 

Koopman operators have been used to characterize observability of a nonlinear system \cite{vaidya_2007,surana_banaszuk_2016}. Yeung et al. formulated the Koopman gramian and showed they can be used to quantify controllability and observability and lend insight for the underlying nonlinear dynamical system \cite{yeung_liu_hodas_2018}. This recent development of the Koopman gramians can advance the imporant and unsolved problem of optimal sensor placement in control theory. The Koopman framework embeds nonlinear dynamics in a linear framework for optimal nonlinear estimation and control \cite{korda_mezic_2018,abraham2017model,arbabi2018data}.  For sensor placement search spaces that are reasonable in size, there are model-based solutions using
optimal experiment design \cite{boyd2004convex,joshi2009sensor}, information
theoretic and Bayesian criteria \cite{caselton1984optimal,krause2008near,lindley1956measure,sebastiani2000maximum,paninski2005asymptotic}. There is a need to develop purely data-driven methods for determining optimal sensor placement. Manohar et al. explored optimized sparse sensor placement for signal reconstruction based on a tailored library of features extracted from training data \cite{manohar2017data}. In \cite{sharma2018transfer}, Sharma et al. extended the transfer operator based approach for optimal sensor placement, providing a probabilistic
metric to gauge coverage under uncertain conditions. Fontanini et al. presented a data driven sensor placement algorithm based on a dynamical systems approach, utilizing the Perron-Frobenius operator \cite{fontanini2016methodology}. Our framework provides a method to determine optimal sensor placement, even in the presence of noisy and temporally sparse data using Koopman operator theory. 

In this paper, we develop an algorithm for optimizing sensor placement from sparsely sampled time-series data.  We use the Koopman observability gramian, developed by Yeung et al. \cite{yeung_liu_hodas_2018}, to maximize the observability of the underlying discrete time nonlinear dynamical system. Section \ref{sec:Koop} introduces the Koopman operator formulation and Section \ref{sec:ObsGram} introduces the notion of a Koopman observability gramian \cite{yeung_liu_hodas_2018}. In Section \ref{sec:ObsPlaceSparse}, we show how to compute the temporally fine-grained Koopman operator from the temporally coarse-grained Koopman operator, which is learned from data that are temporally sparse. In the case of noisy data, a closed form expression for the error in computing the temporally coarse-grained Koopman operator is derived. In Section \ref{sec:OptObs}, we present a novel algorithm for optimal sensor design and placement. Finally, the algorithm is illustrated with a simulation example. 

\section{Koopman Operator Formulation} \label{sec:Koop}
We briefly introduce Koopman operator theory, as we will use it extensively for the sensor placement problem. Consider a discrete time open-loop nonlinear system of the form 
\begin{equation}
\begin{split}
 x_{t+1} &=f(x_t)\\
 y_t &=h(x_t)
\end{split}
 \label{eq:sys}
\end{equation}
with $f:\mathbb{R}^n\rightarrow \mathbb{R}^n$ is analytic and $h$ $\in$ $\mathbb{R}^p$. The Koopman operator of (\ref{eq:sys}), $\mathcal{K}$ : $\mathcal{F}$ $\rightarrow$ $\mathcal{F}$, is a linear operator that acts on observable functions $\psi (x_k)$ and propagates them forward in time as 
\begin{equation}
    \psi (x_{t+1})=\mathcal{K}\psi (x_t).
    \label{eq:KoopEq}
\end{equation}
Here $\mathcal{F}$ is the space of observable functions that is invariant under the action of $\mathcal{K}$. 
\begin{assumption}
Given system (\ref{eq:sys}), we suppose that $y_k=h(x_k)$ $\in$ $\mathcal{F}$ and that $h$ $\in$ span$\left\{\psi_1,\psi_2,...\right\}$.
\end{assumption}
Then the output $y_t$ can be expressed as 
\begin{equation}
    y_t = h(x_t) = W_h\psi(x_t)
    \label{eq:out}
\end{equation}
where the output matrix $W_h$ $\in$ $\mathbb{R}^{p\times n_L}$, $n_L \leq \infty$. We make this strong assumption since the structure of $W_h$ will be manipulated to achieve optimal sensor placement. 

Throughout the paper, we take observable functions which are state-inclusive, i.e.
\begin{equation}
\psi (x) = (x,\varphi(x)) \label{eq:stateIncObs} 
\end{equation}
where $\varphi$ $\in$ $\mathbb{R}^{n_L-n}$ are continuous functions in $\mathcal{F}$.

\section{Koopman Observability Gramian} \label{sec:ObsGram}
The observability matrix of the transformed system may be obtained by showing how the Koopman operator maps initial conditions $x_0$ to $y$ \cite{yeung_liu_hodas_2018}. Using equations (\ref{eq:KoopEq}) and (\ref{eq:out}), we have
\begin{equation}
    y_t = W_h \mathcal{K}^t \psi (x_0)
\end{equation} 
Therefore, $W_h \mathcal{K}^t :$ $\mathbb{R}^{n_L}$ $\rightarrow$ $\mathbb{R}^p$ is the transformation that maps
$\psi (x_0)$  to $y_t$. Given an initial condition $\psi (x_0)$ $\in$ $\mathbb{R}^{n_L}$, the energy of the output $y_t$ is given by
\begin{equation}
\begin{split}
    \Vert y \Vert ^2 & = \sum_{n} <y_t,y_t> \\ 
     & = \sum_{n} \psi (x_0)^\top (\mathcal{K}^t)^\top W_h^\top W_h \mathcal{K}^t \psi (x_0) \\
     & = \sum_{n} \psi (x_0)^\top X_o^{\psi} \psi (x_0)
     \label{eq:normysquared}
\end{split}
\end{equation}
where $<\cdotp,\cdotp>$ represents the inner product and as can be seen in the last equality of (\ref{eq:normysquared}), the Koopman observability gramian is defined as
\begin{equation}
    X_o^{\psi} = \sum_{t=0}^{\infty} (\mathcal{K}^t)^\top W_h^\top W_h \mathcal{K}^t
\end{equation}
and is an $n_L \times n_L$ matrix. The observability gramian can be obtained as a
solution of following matrix Lyapunov equation
\begin{equation*}
    {\cal K}^\top X_o^{\psi} {\cal K} - X_o^{\psi} = -W_h^\top W_h.
\end{equation*}
The Koopman observability gramian quantifies the observability of the function $\psi (x)$. More importantly, when $\psi (x)$
includes observable functions related to the local observability of the underlying nonlinear system (\ref{eq:sys}), the Koopman
observability gramian retains that information \cite{yeung_liu_hodas_2018}. 

\section{Sensor Placement from Temporally Sparse Data} \label{sec:ObsPlaceSparse}
\subsection{\textit{Fine-grained} models from \textit{coarse-grained} models }
We consider the scenario where high-resolution measurements of all genes in a single cell are infrequently sampled.  This is a common scenario when tracking the state of biological, cyber-physical, and social networks.   Exhaustive measurement of every state in the system is expensive (and often manual) and thus can only be performed infrequently.  

Consider the case where the system in (\ref{eq:sys}) is a biomolecular reaction network evolving with unknown governing equations. The precise functional form and parameters of $f$ are typically considered unknown.  In some settings, {\it a priori} knowledge of the biomolecular reaction network can be utilized to bootstrap the modeling problem \cite{yeung2014modeling,yeung2017biophysical} . We consider a data-driven operator theoretic approach, using the method of Koopman briefly introduced in Section \ref{sec:Koop}. 

The discrete time Koopman representation for the system (\ref{eq:sys}) is 
\begin{equation}
    \psi(x_{t+1}) = K\psi(x_t)
\end{equation}
where the matrix $K \in \mathbb{R}^{n_L\times n_L}$ is a finite dimensional approximation of the exact Koopman operator {$\cal K$} and $\psi (x_k) \in \mathbb{R}^{n_L}$. We suppose that full-state measurements are made available for $x_{t}, x_{t+N}$, with enough biological replicates that the {\it temporally coarse-grained} (approximate) Koopman operator is identifiable via the optimization problem 
\[
\min_{K_N} || \Psi(X_f) - K_N\Psi(X_p)||
\]
where 
\begin{equation*}
\begin{aligned}
  \Psi(X_f) &\equiv \begin{bmatrix} \psi(x_{t+N}(\omega_1)) & \hdots & \psi(x_{t+N}(\omega_R)) \end{bmatrix},  \\ \Psi(X_p) & \equiv \begin{bmatrix} \psi(x_{t}(\omega_1)) & \hdots & \psi(x_{t}(\omega_R))\end{bmatrix}.
  \end{aligned}
\end{equation*}
and $\omega_R$ represents the number of replicates. In the presence of sparse and noisy data, \cite{sinha_yeung_2019} showed that the Koopman learning problem can be formulated as a robust optimization problem, which is equivalent to a specific regularized learning problem in which the LASSO penalty parameter corresponds to the upper bound on the noise i.e. the maximum Frobenius norm of the noise. We will suppose, for simplicity of exposition of the technique, that an exact Koopman operator for the coarse-time step mapping $t$ to $t+N$ is either known or obtained directly from data satisfying 
\begin{equation} \label{eq:coarseKoop}
    \psi(x_{t+N}) = K_{N} \psi(x_t).
\end{equation}
Because of linearity of the Koopman operator, we know that the {\it temporally fine-grained} Koopman operator $K$ satisfies
\begin{equation*}
    \psi (x_{t+1}) = K \psi (x_t) 
\end{equation*}
and most importantly, 
\begin{equation} \label{eq:coarsetofine}
    K = K_N^{1/N}.
\end{equation}
When the Koopman observable function includes the state as an element, this relationship allows the recovery of the fine-grained governing equations for $f$ directly from a temporally coarse-grained Koopman operator (and the corresponding data).
To see this, take the state-inclusive observable functions (\ref{eq:stateIncObs}) and partition the Koopman equation accordingly as 
\begin{equation}
\begin{bmatrix}
x_{t+1} \\ \varphi(x_{t+1}) 
\end{bmatrix}
 = \begin{bmatrix} K_{xx} & K_{x\varphi} \\ K_{\varphi x} & K_{\varphi \varphi} \end{bmatrix}\begin{bmatrix}
x_{t} \\ \varphi(x_{t}) 
\end{bmatrix}.
\end{equation}
Since the Koopman operator satisfies 
\begin{equation}
     K \psi(x_t) = \psi(f(x_t))
\end{equation}
for each row, then in particular, the upper half of the Koopman equation satisfies 
\begin{equation}
x_{t+1} = K_{xx} x_t + K_{x\varphi}\varphi(x_t)  = f(x_t).
\end{equation}
This provides a powerful scheme for estimating the governing equations of a fine-grained time-evolving biological process from sparse or coarse-grained temporal measurements, so long as the coarse-grained time measurement is a product of regularly spaced intervals of time in the fine-grained representation.  Again, since RNAseq and proteomic measurements often provide full-state measurements of a network, this in theory can provide sufficient information to recover the Koopman operator, even in the presence of noise \cite{sinha_yeung_2019}.    The key insight and property leveraged is the linearity of the lifted Koopman representation.  One would not be able to obtain the fine-grained dynamics of the governing equations from a coarse grained representation of the governing equations as it is generally not feasible to compute the $n^{\text{th}}$ root of a $n$-layered function composition. 
Specifically, note that the $N$-step map for the underlying governing dynamics of system (\ref{eq:coarseKoop}) is given as 
\begin{equation}
    x_{t+N} = f^{(n)}(x_t) = f \circ f \circ \hdots f (x_t) \equiv f_N(x_t).
\end{equation}
Given an arbitrary nonlinear function $f_N(x_t)$ that is the $N^{\text{th}}$ composition of $f(x_t)$, there is no general way to obtain the underlying $f(x_t)$.  However, by using the Koopman operator lifting framework, we can express the governing equations with linear coordinates, which allows us to consider computing the $N^{\text{th}}$ root to obtain the single-step map from the $N$-step map. 

Although, in general, the matrix root always exists, we note that we may not always obtain the desired fine-grained $K$ from $K_N$ due to there being multiple solutions to matrix roots. Yue et al \cite{yue2016systems} showed that similarly the matrix logarithm raises a concept of system aliasing. They describe the scenario where there might be multiple fine-grained systems which give the same coarse-grained system. In the case that multiple fine-grained Koopman operators exist, our method can be applied to each operator. We can distinguish which operator is the "correct" operator by collecting a few data points at a fine-grained temporal resolution and evaluating the predictive accuracy of the fine-grained Koopman operator models.  

\begin{comment}
\ec{this seems like solid thinking, but the assumption of obtaining a principal matrix root seems too strong. Could you still do anything meaningful if you had another matrix root that satisfied the N-power equation?  If the answer is yes, worth adding this point.  We should discuss this further.}
\end{comment}

In the presence of noise, we approximate the fine-grained discrete time Koopman operator $K$ from the coarse-grained discrete time Koopman operator $K_N$ as
\begin{equation}\begin{aligned}\label{eq:ctofwitherror}
\hat{K} = \hat{K}_N^{(1/N)}&  = (K_N + \epsilon(x))^{1/N}\\ 
& = \sum_{k=0}^\infty \binom{1/N}{k} K_N^{(1/N -k)}\epsilon(x)^k \\
&= \hat{K}_N^{1/N}+\frac{1}{N}K_N^{(1/N - 1)} \epsilon (x) \\ & \qquad 
+ \frac{\frac{1}{N}(\frac{1}{N}-1)}{2!}K_N^{(1/N-2)} \epsilon (x)^2 + ...
\end{aligned}
\end{equation}
where the last equality follows from Newton's generalization of the binomial theorem \cite{liu2010essence}. Here we assume that $\epsilon(x)$ is bounded as in \cite{johnson_yeung_2018} for all $x \in {\cal M}$ $\subseteq$ $\mathbb{R}^n$. A closed form expression of the error term $\epsilon (x)$ is found by noting that
\begin{equation*} 
    \epsilon (x)  = \hat{K}_N - K_N. 
\end{equation*}
Then we have
\begin{equation*} 
\begin{aligned}
    \epsilon (x) \Psi(X_p) & = (\hat{K}_N - K_N) \Psi(X_p)
    & = \hat{\Psi}(X_f) - \Psi(X_f)
\end{aligned}
\end{equation*}
\begin{equation*} 
    \epsilon (x) \Psi(X_p) \Psi(X_p)^\dagger  = (\hat{\Psi}(X_f) - \Psi(X_f)) \Psi(X_p)^\dagger.
\end{equation*}
giving the closed form expression of the error as
\begin{equation} 
    \epsilon (x)  = (\hat{\Psi}(X_f) - \Psi(X_f)) \Psi(X_p)^\dagger.
\end{equation}

Once we obtain the one-step Koopman operator, notice that the Koopman invariant subspace of observable functions is the same as the $N$-step operator.  We suppose, mirroring the scenario presented with transcriptomic and proteomic measurements, that the state is measured completely, in this setting.  The precise coverage of the entire transcriptome and proteome is often a subject of debate, but relative to the spatial sparsity of fluorescence based readout approaches, we shall assume for our purposes that the full state of the network is measured sparsely. 

The state-output equations of the coarse-grained system can then be written as  
\begin{equation}
    \begin{aligned}
        x_{t+N} &= f(x_t) \\ 
        y_t & = x_t
    \end{aligned}
\end{equation}
and thus the corresponding Koopman equation can be written as 
\begin{equation} \label{eq:coarseKoopSys}
\begin{aligned}
    \psi(x_{t+N})& = K_N \psi(x_t)\\ 
    y_t &= P_x \psi(x_t)
    \end{aligned}
\end{equation}
where \[P_x = \begin{bmatrix} I_{n} & 0 \\ 0 & 0 \end{bmatrix}\]
is the projection matrix that extracts the state observable from the vector observable $\psi(x_t).$

\begin{comment}
\ec{This section loses the reader.  Needs a picture showing what we're trying to do and better framing as to the goals of the paper.  Consider dividing into subsections that get us from point A to point B, where point B is the next section.}
\end{comment}

\subsection{Fine-Grained Sensor Placement via Optimal Observability} \label{sec:OptObs}

Often times, it is not physically or economically feasible to measure every state with the resolution specified by a fine-grained temporal model. We seek to develop an algorithm for identifying the design and placement of reporters that maximizes the observability of the underlying nonlinear system, as well as the corresponding Koopman representation.   For this task, we find it convenient to pose this problem using the Koopman gramian as defined in Section \ref{sec:ObsGram}. Specifically, we seek to construct an output observer for the fine-grained dynamical system (\ref{eq:sys}) given full-state sparse temporal measurements at $t$, $t+N$, $t + jN$ in sufficient frequency to recover the temporally coarse-grained Koopman operator $K_N,$ so that it is possible to compute the fine-grained Koopman operator  $K = K_N^{(1/N)}$.  We suppose that the corresponding Koopman representation with output equation is thus written as
\begin{equation} \label{eq:fineKoopSys}
\begin{aligned}
\psi(x_{t+1}) &= K \psi(x_t) \\
y_t &= W_h \psi(x_t).
\end{aligned}
\end{equation}
We seek to maximize the output energy $||y_t||^2$ for an initial condition $x_0$ at a time instant $t$ i.e. solve the nonlinear optimization problem 
\begin{equation} \label{eq:NPoptProb}
    \max_{h(x) \in {\cal L}^2\left({\cal M}\right)}   ||y(t_j)||^2
\end{equation}
for all initial conditions $x_0$ with $||x_0|| \leq 1$.  This is an optimization problem of a nonlinear function space (i.e. an uncountably infinite dimensional space) and is generally intractable. However, if we were to find a basis for $h(x)$, we could express the problem in terms of a linear combination of the basis functions, which would yield a convex formulation of the problem.  This is precisely what we can do using the spectral properties of the Koopman operator representation. 
\begin{comment}
Here we make the assumption that $f$ is at least Lyapunov stable and $K$ is at least marginally stable.  \ec{why do you need these assumptions? explain... are they really necessary?}
\end{comment}
Following the formulation given in Section \ref{sec:ObsGram}, the system in (\ref{eq:fineKoopSys}) has Koopman observability gramian
\begin{equation} \label{eq:ObsGramFine}
    X_{o,f}^\psi = \sum_{j=0}^{t_N} (K^{j})^\top W_h^\top W_h (K^{j})
\end{equation}
where the subscript $f$ is used to distinguish the fine-grained system from coarse-grained.

We want to identify the optimal sensor placement that informs the design of optimal observers. Utilizing the Koopman observability gramian, $X_{o,f}$, as defined in (\ref{eq:ObsGramFine}), the output energy of system (\ref{eq:fineKoopSys}) is written as
\begin{equation} \label{eq:outputNorm}
    ||y_{t_N}||^2 = \sum_{j=0}^{t_N} \psi (x_0)^\top (K^{j})^\top W_h^\top W_h (K^{j}) \psi (x_0).
\end{equation}
Our goal is to now maximize the output energy (\ref{eq:outputNorm}) of the lifted system up at time $t$ with the output matrix $W_h$ as the decision variable. If the output energy of the lifted system is maximized, then by proxy the output energy of the original nonlinear system is maximized.

For the purposes of this paper, we will suppose that we construct an observable function basis that results in a diagonalizable Koopman operator. The subsequent presentation can be generalized for Koopman operators that only admit a Jordan decomposition, but for simplicity of exposition, we consider the case of the diagonalizable Koopman operator. 
\begin{assumption}
We suppose that $\psi(x)$ and $K$ are provided or trained during the learning process to admit a diagonalizable $K$. 
\end{assumption}
Thus, an eigendecomposition of $K$ gives
\begin{equation*}
    KV = V\Lambda
\end{equation*}
where $V$ is an $n_L \times n_L$ matrix of eigenvectors. The $n_L \times n_L$ matrix $\Lambda$ is a diagonal matrix whose components are the eigenvalues $\lambda$ of the Koopman operator, $K$. The eigenfunctions of $K$ are then written as
\begin{equation*}
    \phi (x_0) = V^{-1}\psi(x_0).
\end{equation*}
where $\phi \in \mathbb{R}^{n_L}$. Since (\ref{eq:outputNorm}) has a symmetric form, let us deal with the right half of this equation. We have that
\begin{equation*}
\begin{aligned}
    W_hK^j\psi(x_0) &= W_h V\Lambda^j V^{-1}\psi(x_0) \\ 
    &= W_h V\Lambda^j V^{-1} V \phi (x_0) \\
    &= W_h V\Lambda^j \phi(x_0).
\end{aligned}
\end{equation*}
 The output energy can now be written in terms of the Koopman eigenfunctions as 
\begin{equation} 
    ||y_{t_N}||^2 = \sum_{j=0}^{t_N} \bigg[\phi(x_0)^\top\Lambda^jV^\top W_h^\top W_hV\Lambda^j\phi(x_0) \bigg]  
\end{equation}
The optimization problem (\ref{eq:NPoptProb}) can now be formulated as
\begin{equation} \label{eq:optProb}
    \mathcal{J} = \max_{W_h} \sum_{j=0}^{t_N} \bigg[\phi(x_0)^\top\Lambda^jV^\top W_h^\top W_hV\Lambda^j\phi(x_0) \bigg]
\end{equation}
with $|| W_h^\top W_h ||_2 \leq C$. The upper bound $C$ would vary between biological experiments and should be identified directly from data. 

By picking out the $p$ $(\leq n_L)$ most observable modes of the system such that we can ensure the collection of measurements which correspond to maximal energy. If we define $W_h$ as
\begin{equation*}
    W_h \triangleq \begin{bmatrix} I_{p\times p} & 0 \end{bmatrix}  V^{-1}
\end{equation*}
the argument of (\ref{eq:optProb}) becomes 
\small
\begin{equation*}
\begin{aligned}
    & \sum_{j=0}^{t_N} \left( \phi(x_0)^\top\Lambda^jV^\top(V^{-1})^\top \begin{bmatrix} I_{p\times p}  \\ 0  \end{bmatrix}  \begin{bmatrix} I_{p\times p}  & 0  \end{bmatrix} V^{-1}V\Lambda^j\phi(x_0) \right)  \\
    & \qquad = \sum_{j=0}^{t_N} \left(\phi(x_0)^\top  diag(\lambda_1^{2j},\lambda_2^{2j},
    ...,\lambda_p^{2j},0,...,0) \phi(x_0) \right)
\end{aligned}
\end{equation*}
\normalsize
where $\lambda_1$ through $\lambda_p$ are the $p$ maximum eigenvalues of $K$. The maximum output energy comes from a choice of $W_h$ that depends on the eigenvectors of the Koopman operator.

\subsubsection{Example (Circadian oscillator)}
To illustrate our sensor placement algorithm, we consider a model of a circadian oscillator, see Vilar et al. \cite{vilar2002mechanisms}, that involves an activator $A$ and a repressor $R$. Both $A$ and $R$ are transcribed into $mRNA$ and subsequently translated into protein. Since $A$ can bind to both $A$ and $R$ promoters, it increases their transcription rates. $R$ acts as a negative element by hindering $A$. The deterministic dynamics are given by the following reaction rate equations

\begin{equation} \label{eq:circadianModel}
\begin{aligned}
    \dot{D}_A  &= \theta_AD_A^{'} - \gamma_AD_AA \\
    \dot{D}_R  &= \theta_RD_R^{'} - \gamma_RD_RA \\
    \dot{D}_A^{'}  &= \gamma_AD_AA - \theta_AD_A^{'} \\
    \dot{D}_R^{'}  &= \gamma_RD_RA - \theta_RD_R^{'} \\
    \dot{M}_A &= \alpha_A^{'}D_A^{'} + \alpha_AD_A - \delta_{MA}M_A \\
    \dot{A}  &= \beta_AM_A + \theta_AD_A^{'} + \theta_RD_R^{'} \\ & - A(\gamma_AD_AA + \gamma_RD_R + \gamma_CR + \delta_R) \\
    \dot{M}_R  &= \alpha_R^{'}D_R^{'} + \alpha_RD_R - \delta_{MR}M_R \\
    \dot{R}  &= \beta_RM_R - \gamma_CAR  + \delta_AC - \delta_RR \\
    \dot{C}  &= \gamma_CAR - \delta_AC.
\end{aligned}
\end{equation}

Extended dynamic mode decomposition (EDMD) \cite{williams_kevrekidis_rowley_2015} is used to compute the finite-dimensional approximation of the Koopman operator, $K_N$, for a coarse time step. A dictionary of state-inclusive observable functions, $\Psi$, is constructed using up to second-order polynomials.  Often in biological systems, Hill function type nonlinearities appear in the dynamics. Even in these cases, the dictionary of polynomial functions should capture the dynamics well, according to the Weierstrass Approximation theorem, which states that any continuous function on a closed and bounded interval can be uniformly approximated on that interval by polynomials. \cite{stone}. Therefore, as long as the non-polynomial linearity is continuous, we expect that this dictionary of polynomials will result in accurate predictions, although the representation may not be as low dimensional as a representation drawn from a more efficient encoding \cite{lusch2018deep,yeung2017learning}.

\begin{figure}[t]
    \centering
    \includegraphics[width=0.85\columnwidth]{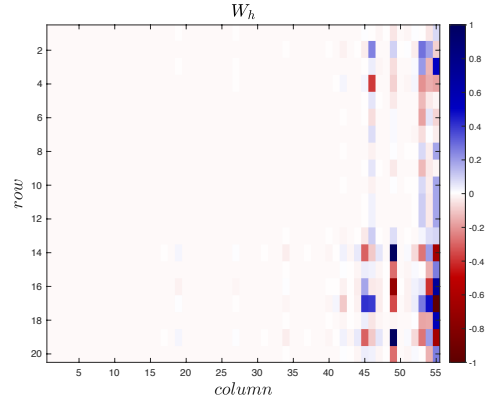}
    \caption{Sparse structure of the output matrix $W_h$ with $p=20$ for the circadian oscillator simulation.}
    \label{fig:WhStructure2}
\end{figure}

\begin{figure}[t]
    \centering
    \includegraphics[width=0.61\columnwidth]{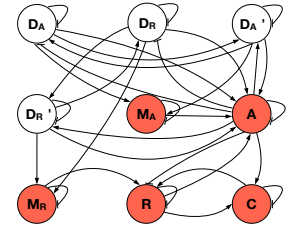}
    \caption{Network architecture of the circadian oscillator model in (\ref{eq:circadianModel}). Arrows indicate activation, while bars indicate repression or degradation. Highlighted in red are the states which have the most active dynamics in the observable coordinates. Note that these active states were taken from the single initial condition used to produce figure \ref{fig:WhStructure2}.}
    \label{fig:NetArch}
\end{figure}

\begin{figure}[t]
    \centering
    \includegraphics[width=0.85\columnwidth]{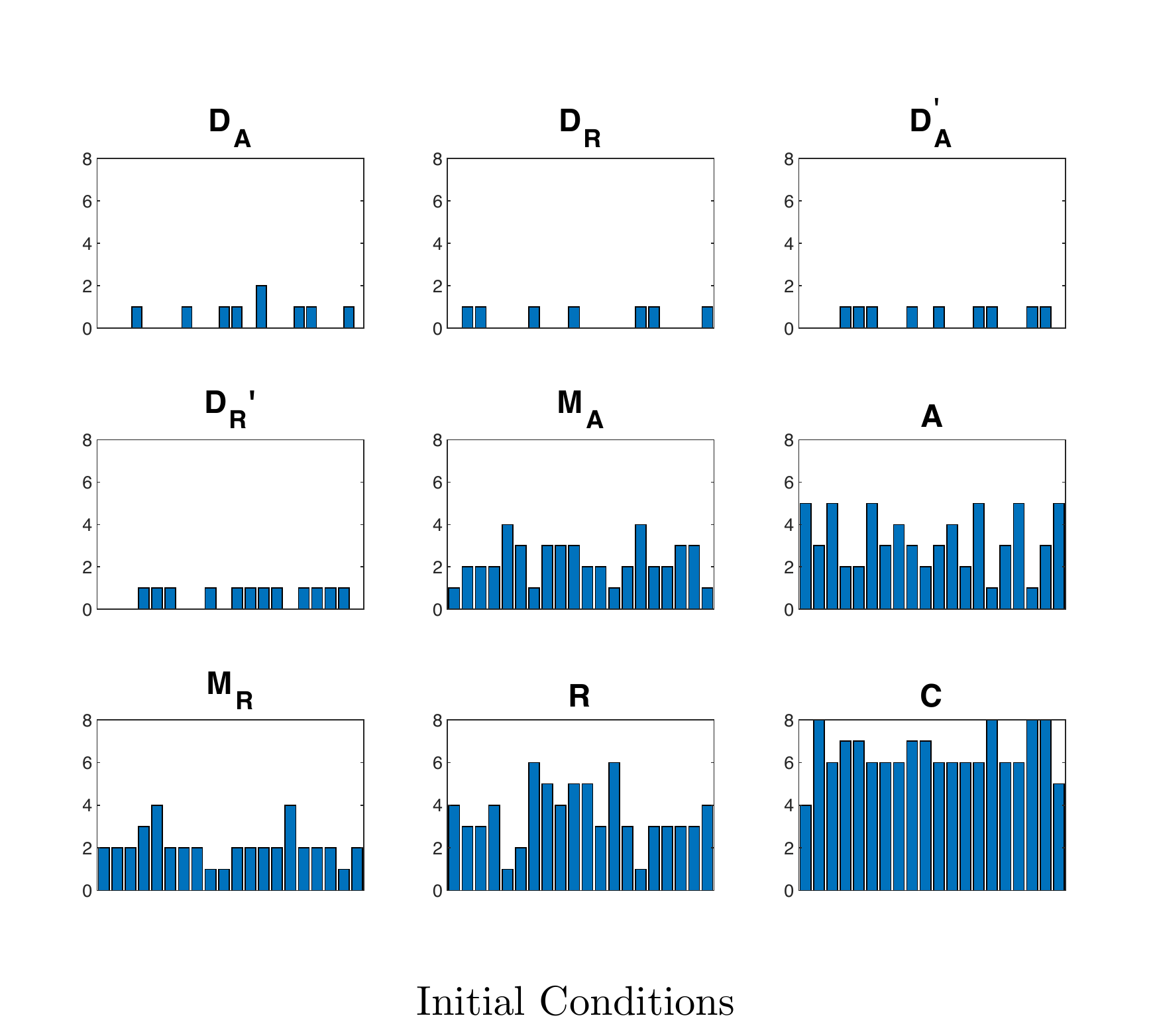}
    \caption{Histogram showing the frequency of a state of (\ref{eq:circadianModel}) being in the 10 most active states of the observables over 20 different initial conditions.}
    \label{fig:actDyn}
\end{figure}

In this example, initial conditions were chosen such that the trajectories converge to a limit cycle. From the coarse-grained Koopman operator obtained from simulation data, the fine-grained Koopman operator, $K$, is computed using the scheme outlined in Section \ref{sec:ObsPlaceSparse}. Solving the optimization problem (\ref{eq:optProb}), we can identify the optimal sensor placement. Choosing $p$, the number of rows in $W_h$, to be $p=20$, we get the output matrix structure as seen in figure \ref{fig:WhStructure2}. A total of 55 observable functions were used which correspondingly sets the number of columns in the output matrix, $W_h$. The output matrix has a sparse structure with most elements of the matrix nearly zero. Using the criteria that the 1-norm of the columns of $W_h$ determine the most \textit{active} states of the observable coordinates, we can determine optimal sensor placement. Using this criteria, the most active dynamics are $M_AC, M_RC, AC, R^2, RC, \text{and }C^2$ for a single initial condition where the trajectories converge to limit cycles. Figure \ref{fig:NetArch} shows the entire network architecture of the circadian oscillator. The states highlighted in red are the active states and correspondingly are where the algorithm would dictate sensors should be placed. Figure \ref{fig:actDyn} shows how frequently a state appears as an active state in the observable coordinates over 20 different initial conditions. 

From this analysis, a nonlinear observable can be designed. For example, the state $C^2$ is highly active in the observable basis, therefore a nonlinear observer can be designed where a $C$ molecule binds with another $C$ molecule and integrated to obtain the output. We can then use this observer (and other observers) to act as a reporter for the cell state. This can enable rapid experimentation in synthetic biology since there would no longer be a need to collect expensive full state proteomics and transcriptomics data at a low temporal resolution. We can collect partial state measurements from states of interest at a high temporal resolution. The Koopman method thus can identify critical genes that serve as cell state biomarkers. These biomarkers provide a link between internal dynamics and observed phenotypes.  
 
\section{Conclusion}
In this work, we developed an algorithm for optimal sensor placement from sparsely sampled time-series data of discrete time nonlinear systems. The optimal sensor placement algorithm was formulated as maximizing the observability of a dynamical system or genetic network in the Koopman lifted space. We compute the temporally \textit{fine-grained} Koopman operator from the temporally \textit{coarse-grained} Koopman operator, the latter of which is identified directly from sparse biological data. In the case of noisy data, a closed form expression for the error in the coarse-grained Koopman operator is derived. Finally, we have illustrated the optimal sensor placement method on a simulation example of a circadian oscillator. This method can be utilized in the context of developing bacterial sensors where the design of a library of promoters is now informed by the sensor placement algorithm. 

\section*{Acknowledgements}
The authors would like to thank Professor Igor Mezic for insightful discussions.  
This material is based on work supported by DARPA and AFRL under contract number DEAC0576RL01830. Any opinions, findings, conclusions, or recommendations expressed in this material are those of the authors and do not necessarily reflect the views of the Defense Advanced Research Project Agency, the Department of Defense, or the United States government. 

\bibliographystyle{abbrv}
\bibliography{main}

\end{document}